\documentclass[11pt, letterpaper]{article}
\usepackage{amssymb}
\usepackage{fullpage}
\usepackage[noadjust]{cite}
\usepackage[noend]{algpseudocode}
\usepackage{times}
\title{Arithmetic Progression Hypergraphs: \\ Examining the Second Moment Method}

\author{Michael Mitzenmacher\thanks{Harvard University, School of Engineering and Applied Sciences. Supported in part by NSF grants CCF-1320231, CNS-1228598, IIS-0964473, and CCF-0915922. Part of this work was done while visiting Microsoft Research, New England.} \\ michaelm@eecs.harvard.edu }

\date{}

\usepackage{amsfonts,amsmath,color,float,graphicx,verbatim,url, amsthm}

\usepackage{algorithm}

\usepackage{enumerate}

\newtheorem{theorem}{Theorem}
\newtheorem{lemma}[theorem]{Lemma}

\newtheorem{corollary}[theorem]{Corollary}

\newtheorem{definition}[theorem]{Definition}

\newcommand{\xor}{{\oplus}}

\newcommand{\E}{\mathbb{E}}

\newcommand{\eat}[1]{}

\usepackage{soul}

\usepackage{color}

\newcommand{\knae}{$k$-NAE-SAT}
\newcommand{\knaespace}{$k$-NAE-SAT }
\newcommand{\kapnae}{$k$-AP-NAE-SAT}
\newcommand{\kapnaespace}{$k$-AP-NAE-SAT }

\newcommand{\ksatspace}{$k$-SAT }
\newcommand{\kapsat}{$k$-AP-SAT}
\newcommand{\kapsatspace}{$k$-AP-SAT }

\begin{document}
\pagenumbering{gobble}
\maketitle
\begin{abstract}
In many data structure settings, it has been shown that using ``double
hashing'' in place of standard hashing, by which we mean choosing
multiple hash values according to an arithmetic progression instead of
choosing each hash value independently, has asymptotically negligible
difference in performance.  We attempt to extend these ideas beyond
data structure settings by considering how threshold arguments based
on second moment methods can be generalized to ``arithmetic
progression'' versions of problems.  With this motivation, we define a novel
``quasi-random'' hypergraph model, random arithmetic progression (AP) hypergraphs,
which is based on edges that form arithmetic
progressions and unifies many previous problems.  Our main result is
to show that second moment arguments for 3-NAE-SAT and 
2-coloring of 3-regular hypergraphs extend to the double hashing
setting.  We leave several open problems related to these quasi-random
hypergraphs and the thresholds of associated problem variations.
\end{abstract}

\newpage
\pagenumbering{arabic}

\section{Introduction}
\subsection{Arithmetic Progression Problems}
For several multiple-choice hashing data structures, it has been shown
that using ``double hashing'' instead of uniform hashing yields
essentially no difference in performance, in both the asymptotic and
practical senses of performance.  For example, consider the balanced
allocation setting \cite{azar1999balanced} where $n$ balls are placed into $n$ bins, but
instead of placing them randomly according to a single hash function
(assumed to be perfectly random), each ball is placed sequentially in
the least loaded of $d$ choices, each obtained independently by a hash
function, for some constant integer $d > 2$.  The asymptotic fraction
of bins of each constant load can be found by fluid limit differential
equations \cite{mitzenmacher1999studying}, and the maximum load is known to be $\log \log n/\log d +
O(1)$ with high probability \cite{azar1999balanced}.  Now suppose the $d$ choices are chosen
in a fashion reminiscent of double hashing, in the following way.  Let
us assume that $n$ is prime, and each ball is hashed twice, to obtain
$a$ chosen uniformly from $[0,n-1]$ and $b$ chosen uniformly from
$[1,n-1]$.  The $d$ choices for a ball are then given by bins $a, a+b,
a+2b, \ldots, a+(d-1)b$ modulo $n$.  That is, the $d$ choices are
constrained to form an arithmetic progression; as such, only two
random numbers modulo $n$ are chosen to determine the choices, instead
of $d$ random numbers.  It has been shown that, even with this more
limited randomness, not only does the maximum load remain $\log \log
n/\log d + O(1)$ with high probability, but that the asymptotic
fraction of bins of each constant load is the same as when the $d$
choices are all perfectly random \cite{mitzenmacher2014balanced,mitzenmacher2016more}.  
Similar results have been found for other data structures that use multiple hash functions,
including Bloom filters \cite{kirsch2008less} and cuckoo hashing \cite{mitzenmacher2012peeling,leconte2013double}.

In this paper, we expand on this theme to consider whether similar
phenomena occur in ``multiple-choice'' settings outside of hash-based
data structures.  Specifically, we attempt to generalize threshold
arguments for variations of random constraint problems found using the
second moment method, such as variations of \knaespace and 2-coloring random
$k$-uniform hypergraphs.
Formalizing these variations lead us to propose a novel ``quasi-random''
hypergraph model that we believe is worthy of future study in its own
right; we propose several open questions related to this graph model.

To review, the standard \knaespace problem is the following: given a
collection of variables $\{x_1,\ldots,x_n\}$, and a collection of
clauses where each clause consists of $k$ literals on these variables
(that is, $k$ variables or negations of variables), can we assign
true/false values to the variables so that each clause has at least
one true and at least one false literal.  For the random \knaespace
problem, $m$ clauses are chosen uniformly at random from the
collection of all clauses.  (There are variations depending on whether
literals for each clause are chosen with or without replacement; the
choice of such details can simplify analysis, but don't affect the
asymptotic results we consider here.)

Our proposed variation, \kapnae, adds the following additional
requirements: the number of variables $n$ is prime, and the $k$
variables in each clause must form an arithmetic progression $a, a+b,
a+2b, \ldots, a+(k-1)b$ modulo $n$.  In the random variation, each
clause is chosen independently and uniformly from the $2^kn^2$
possible clauses; that is, the $k$ signs for the literals are chosen
independently and uniformly at random, and we choose values for $a$
and $b$ modulo $n$ independently and uniformly at random to choose the
variables within a clause.    Our
primary result is that a second moment method argument of Achlioptas
and Moore for random \knaespace can be applied to random \kapnaespace to obtain
a lower bound threshold result for the case where $k=3$
\cite{achlioptas2006random}. 

The \knaespace problem is similar to the hypergraph $2$-coloring problem of
$k$-uniform hypergraphs: given a collection of $m$ hyperedges each of
$k$ vertices, can the vertices be colored with two colors so that each
edge has a vertex of each color. Here we consider the variation of
2-coloring hypergraphs, where each hyperedge consists of vertices that
form an arithmetic progression.  We again show that a second moment
argument due to Achlioptas and Moore generalizes to this setting when
$k=3$ \cite{achlioptas2002}.

We describe the current impediments to extending these results beyond
$k=3$, and make corresponding conjectures.  More generally, we
describe a ``quasi-random'' hypergraph model where edges are
constrained to form an arithmetic progression that captures these and
many other problems.  We suggest several open questions related to
this quasi-random hypergraph model.  

\subsection{Model and Motivation}  
As described above, we consider variations of random \knaespace where the
variables within a clause form an arithmetic progression, which may
appear to be a somewhat unusual object.  We offer multiple motivations
for why this is an interesting problem to study.

Satisfiability problems are a particularly natural problem to
consider in this context, as random $k$-SAT and variations have been the subject of
wide study, with a great deal of work focusing on the thresholds
for satisfiability (see, e.g.,
\cite{achlioptas2006random,coja2012catching,ding2015proof,friedgut1999sharp}
and references therein).  In particular, there has been some belief
that random problems at or near the threshold boundary may be where
``hard instances'' lie for NP-complete variations such as $k$-SAT
\cite{cheeseman1991really,monasson1999determining}.  It therefore
seems interesting that there may be random $k$-SAT variations utilizing
much less randomness but that seem to have similar threshold properties;
such variations may prove useful for understanding the complexity of 
random $k$-SAT problems or similar problems, or for testing algorithms
for such problems.  Indeed, the added structure of arithmetic progressions
may potentially make such problems easier to study in some circumstances.  

In considering problems with thresholds, the second moment method is a natural
line of attack, given that it led the way to key threshold results for many
problems relatively recently, and is reasonably well understood.  We therefore
turn to the seminal results of Achlioptas and Moore \cite{achlioptas2002,achlioptas2006random}, which   
provided a spark to the area.  

While not specifically a hashing problem, random $k$-SAT problems are
similar (and closely connected to) hashing problems; see
e.g. \cite{dietzfelbinger2010tight}.  As previously noted, there are
various settings where using double hashing has essentially the same
behavior as hashing with fully random multiple choices.  Most of these
hashing problems have natural interpretations as problems on random
hypergraphs, by viewing elements as hyperedges and buckets as vertices
in a hypergraph.  For example, the question of whether there is an
assignment of elements to buckets in cuckoo hashing is equivalent to
the question of whether there is an orientation of hyperedges in a
random hypergraph that does not ``overload'' any vertex by orienting
too many edges toward it.  It therefore seems natural to ask whether
other hypergraph problems behave similarly when using edges based on
arithmetic progressions instead of fully random edges.

The double hashing variation of these hashing schemes correspond to
the following variations of random hypergraph models, which appear
implicit in previous works but do not seem to have been described.  As
mentioned previously, in what follows for convenience we allow for
$n^2$ arithmetic progressions on the numbers modulo $n$ for prime $n >
2$; we include the trivial progressions $(a,a,\ldots,a)$ where all
elements are the same, and we count $(a,a+x,a+2x,\ldots,b)$ and
$(b,b-x,b-2x,\ldots,a)$ as distinct progressions for $0 \leq a < n$,
$0 < x < n$, and $b=a+(k-1)x$. 

\begin{definition}
For a prime number $n$ and $k < n$, let ${H}_k^{AP}(n,p)$ be a random $k$-uniform hypergraph on $n$ vertices where 
where each of the $n^2$ possible edges given by $(a,a+x,a+2x,\ldots,a+(k-1)x)$ with $0 \leq a < n$ and
$0 \leq x < n$ is included with probability $p$. 

Similarly, let ${H}_k^{AP}(n,m)$ be a random $k$-uniform hypergraph
on $n$ vertices with $m$ edges where the edges are determined by
selecting independently and uniformly at random, with replacement,
each edge from all $n^2$ possible arithmetic progressions.
\end{definition}

We may also use the phrasing ``a hypergraph chosen from ${H}_k^{AP}(n,m)$''
(or ${H}_k^{AP}(n,p)$) where the meaning is clear.  We suggest referring
to these types of hypergraphs as {\em random arithmetic progression hypergraphs},
or more succinctly {\em random AP hypergraphs}.

One can vary this definition to disallow hyperedges with repeated vertices
if needed, or to disallow multi-hyperedges, or to have non-regular edges.  Also, we note tht we focus on $n$ prime
for convenience, to avoid asymmetries when $b$ is not relatively prime to $n$.
(In many settings, this issue may ignored without affecting the asymptotics, or similarly $b$ may
instead be restricted to be a randomly chosen number relatively prime to $n$.)

The $H_k^{AP}(n,p)$ model is a natural ``quasi-random'' graph
model that has not, to our knowledge, been studied previously.  Previous
results for hashing schemes suggest it is very like the standard
random hypergraph model $H_k^{AP}(n,p)$ in useful ways (at least
from the algorithmic perspective).  We believe this graph model is
worthy of further investigation, as we discuss below.   

We correspondingly define variations of the $k$-SAT problem with the addition 
of an ``AP'' to denote that the variables must be chosen to form an arithmetic
progression.  Hence instead of \ksatspace and \knae, we may have \kapsatspace and \kapnae.

\subsection{Result Summary and Some Conjectures}

Reviewing previous second moment arguments for threshold
behaviors, several of them rely on determining the distribution of the number of 
monochromatic collections of $k$ variables for a random coloring of the variables 
with two colors (such as ``true'' of ``false'' in the $k$-SAT settings).  
In the variations we consider where the variables in our constraints are 
governed by an arithmetic progression, the relevant question to consider is
the distribution of the number of monochromatic collections of $k$ variables,
when the variables are in an arithmetic progression.  For general $k$, this is
a very challenging problem \cite{bhattacharya2016upper,lu2012monochromatic,warnke2016upper}, but for the case of $k=3$, the distributions
are the same as in the case of complete randomness;  we discuss and prove this below.  This simple fact provides our
main results.  

These results, however, provides some fodder for additional conjectures.
\begin{enumerate}
\item Can the second moment method arguments of \cite{achlioptas2002,achlioptas2006random} for \knae,
or other second moment arguments (such as in \cite{achlioptas2004threshold,coja2012catching}), be made to apply
to \kapnaespace to obtain corresponding threshold results for all $k \geq 3$?  (The same questions can be asked
for hypergraph 2-colorability.)
\item There has been further progress on the actual thresholds for \knaespace and \ksatspace using ``belief propagation''
based analysis (e.g, \cite{ding2015proof,ding2016satisfiability}).  Can these techniques be used and these
results be shown to also hold in the settings of \kapnaespace and \kapsat?
\end{enumerate}
We conjecture all of these questions have positive answers.
We discuss some of the challenges to proving the first conjecture below.

\section{Second Moment Argument for \kapnae}

\subsection{Problem Statement}

We consider \kapnae, following closely the argument of
Achlioptas and Moore \cite{achlioptas2006random}.  We consider formulas $F$ with
variables $x_1,x_2,\ldots,x_n$ and $m = rn$ clauses.  We use the term
literal to refer to a variable or a negated variable.  Fitting with our previous
discussion, 
assume $n$ is a prime and $F$ is determined by choosing each clause
independently and uniformly at random from all clauses with $k$
literals with replacement, where the $k$ literals are required to be one of the $n^2$ possible arithmetic progressions modulo $n$.  
(The analysis is easily shown to be essentially the same if the clauses
cannot include the trivial clauses and/or are chosen without replacement;
see the discussion for instance in Section 4.1 of \cite{achlioptas2006random} showing
such variations affect the threshold by $o(1)$ terms.)  
We aim to determine as tightly as possible bounds for the threshold for satisfiability;
ideally, we would find a constant $r^*$ so that for any constant $\epsilon > 0$, for all $r > (r^*+\epsilon)$ 
with high probability a randomly selected formula $F$ would not be satisfiable, while for 
all $r < (r^*-\epsilon)$, 
with high probability a randomly selected formula $F$ would be satisfiable.
Our results
will hold only for $k=3$, but we write the argument for general $k$
and later explain why $k=3$ is required.

\subsection{Adapting the Second Moment Argument}
Let $X= X(F)$ be the number of NAE-satisfying truth assignments for a
formula $F$; that is, the truth assignment satisfies that for every
clause in $F$, there is at least one satisfied and at least one
unsatisfied literal.  We apply the second moment method as follows.  
The first moment is simple to compute, and indeed the fact that the clauses
are constrained to be arithmetic progressions does not affect the calculation.
$$\E[X] = 2^n \left (1-2^{1-k} \right)^m,$$
since the probability a clause NAE-satisfies a randomly chosen clause is $1-2^{1-k}$.
(In particular, we note the following well-known result for purely random clauses hold here in the
setting of random AP clauses as well:  if $m = rn$ and $k=3$, then for $r > \log_{4/3} 2 \approx 2.41$,
$$\E[X] = 2^n (3/4)^{rn} = o(1),$$
so that first order methods give an upper bound on the satisfiability threshold.)

To compute $\E[X^2]$, we consider all possible $4^n$ ordered pairs of
assignments $(\sigma,\tau)$.  We recall from \cite{achlioptas2006random} that the
probability that both $\sigma$ and $\tau$ NAE-satisfy a randomly
chosen clause $c$ (from the collection of {\em all} possible clauses with
$k$ literals) depends only on the overlap between the assignments.
That is, if $\sigma$ and $\tau$ agree on $z = \alpha n$ variables,
then by inclusion-exclusion we find
$$\Pr[\sigma \mbox{ and } \tau \mbox{ NAE-satisfy } c] = 1 - 2^{2-k}+2^{1-k}\left ( \alpha^k + (1-\alpha)^k\right ).$$
This is because for a clause $c$ to {\em not} be satisfied by both $\sigma$ and $\tau$, the literals of $c$ must
evaluate to all true or all false in $\sigma$ and $\tau$.  One way of expressing this is that if we consider
truth assignments as bit strings in the natural way (with true being 1 and false being 0), and then consider
$\sigma \xor \tau$ as a bit string, a clause $c$ is not satisfied by both $\sigma$ and $\tau$ if and only if the variables in the clause $c$
are {\em monochromatic} in $\sigma \xor \tau$;  that is, they must all be 0 or all 1.  
The probability of a randomly chosen clause $c$ being monochromatic in this manner
is $\alpha^k + (1-\alpha)^k$.  (Technically, this is the probability if the variables are not constrained to be distinct;
as mentioned, the difference between allowing or not allowing repeated variables does not affect the overall threshold argument.)
This expression for the probability $\sigma$ and $\tau$ NAE-satisfy $c$ forms the basis for their calculations regarding bounding the ratio $\E[X^2]/\E[X]^2$ for the remainder of their second moment argument.  

Let us consider $\E[X^2]$ now where $c'$ is a randomly chosen clause
constrained to be an arithmetic progression.  
As before, we consider all possible $4^n$ ordered pairs of
assignments $(\sigma,\tau)$.  Let $\beta(\sigma,\tau)$ be the fraction of 
arithmetic progressions of length $k$ in the bit string for
$\sigma \xor \tau$ that are monochromatic.  
The corresponding inclusion-exclusion for 
this case give us now:
$$\Pr[\sigma \mbox{ and } \tau \mbox{ NAE-satisfy } c'] = 1 - 2^{2-k}+2^{1-k}\beta(\sigma,\tau).$$

For general $k$, $\beta(\sigma,\tau)$ depends on the actual choices of
$\sigma$ and $\tau$.  For the case of $k=3$, however, we have that if
$\sigma$ and $\tau$ agree on $z = \alpha n$ variables, then
$\beta(\sigma,\tau)$ is a function of $\alpha$, and in fact equals $\alpha^3+(1-\alpha)^3$.  In particular, we have the following known result:
\begin{lemma}
\label{lem:mono}
For a prime $n$, a two-coloring of $Z_n$ where the two color classes consist of $\alpha n$ numbers
and $(1-\alpha)n$ numbers has $(1-3\alpha+3\alpha^2)n^2 = (\alpha^3 + (1-\alpha)^3)n^2$ monochromatic arithmetic sequence of length three.
\end{lemma}
Although this result is known in the literature \cite{lu2012monochromatic}, for completeness we provide a proof (suggested to us by Yufei Zhao)
in the appendix.  

Because the probability a randomly chosen
progression of length three is monochromatic is also $(\alpha^3+(1-\alpha)^3)$,
for the second moment calculations of \cite{achlioptas2006random}, the results when clause variables are constrained to be
arithmetic progression of length three remains the same as if the clause variables were randomly chosen.  In particular, 
if we let 
$$f(\alpha) = \frac{1}{2}+\frac{1}{4}(\alpha^3 + (1-\alpha)^3),$$
then following their work, we have
$$\E[X]^2 = 2f(\alpha)^{rn},$$
and (letting $H(x)$ be the binary entropy of $x$)
\begin{eqnarray*}
\E[X^2] & = & \sum_{z=0}^n {n \choose z} f(z/n)^m \\
        & \leq & (n+1) \max_\alpha \left ( {2^{H(\alpha)} f(\alpha)^r} \right )^n \\
        & \leq & C \E[X]^2
\end{eqnarray*}
for some constant $C$.  (We note the second line, bounding $\E[X^2]$ by $C\E[X]^2$, requires
a significant calculation argument, and refer the reader to \cite{achlioptas2006random}.)

We have the following corollary, which
corresponds to the significant work of Lemma 2 and Theorem 5 of \cite{achlioptas2006random}:  
\begin{corollary}
\label{cor:k3kapnae}
For $k=3$, when $r < 1.5$ a random \kapnaespace formula is satisfiable with probability bounded below by some constant.  
\end{corollary}

\subsection{Remarks on the Result}

We would like to further say that Corollary~\ref{cor:k3kapnae} implies
that a random \kapnaespace has a solution with probability $1-o(1)$.  In
\cite{achlioptas2006random}, the authors rely on the results of
Friedgut \cite{friedgut1999sharp} to claim that the result with
uniformly positive probability can be boosted to with high
probability.  Specifically, Friedgut's work is in a line of results
that show for many problems on random graphs and hypergraphs that if
there is a threshold, it must be a ``sharp threshold'' instead of a
``coarse threshold''
\cite{bourgain1997influences,friedgut1996every,friedgut1998boolean,friedgut1999sharp,friedgut2005hunting}.
In particular, this implies that if the second moment results yield a
constant lower bound on the existence of solution, then in fact a
solution exists with high probability.
While these results intuitively should also apply 
as well to the settings of $H_k^{AP}(n,p)$ and $H_k^{AP}(n,m)$,
it is not immediate that they do so;  in particular, these quasi-random
hypergraphs do not have the symmetry properties that fully random hypergraphs
do, which are used in many of the existing proofs.  (Some results here do
not require complete symmetry, but have other non-trivial requirements.)  

We conjecture that the ``sharp threshold'' framework of Friedgut (and
others) can be generalized to the quasi-random hypergraph models
$H_k^{AP}(n,p)$ and $H_k^{AP}(n,m)$; however, we leave this as an open
question.

Naturally, we would also like the result to hold for all constant $k$,
not just $k=3$.  Recall that the second moment argument depends on 
$$\Pr[\sigma \mbox{ and } \tau \mbox{ NAE-satisfy } c'] = 1 - 2^{2-k}+2^{1-k}\beta(\sigma,\tau),$$
where $\beta(\sigma,\tau)$ is the fraction of 
arithmetic progressions of length $k$ in the bit string for
$\sigma \xor \tau$ that are monochromatic.  
For $k=3$, we found this fraction was
$$f_k(\alpha) = 1 - 2^{2-k}+2^{1-k}(\alpha^k + (1-\alpha)^k)$$
for every $\sigma$ and $\tau$.
Unfortunately, for $k > 3$, the fraction of arithmetic progressions
of length $k$ that monochromatic varies with $\sigma$ and $\tau$.

One might hope that, say for randomly chosen $\sigma$ and $\tau$, the
fraction of monochromatic arithmetic progressions in $\sigma \xor
\tau$ would be well concentrated around $f_k(\alpha)$, and this would be
sufficient to bound the second moment and apply the second moment
method.  Chernoff-like concentration bounds for this setting do exist
\cite{bhattacharya2016upper,warnke2016upper}, but it is also known that the number of
arithmetic progressions can differ significantly from the mean (see,
e.g., \cite{lu2012monochromatic}).  The second moment argument of
Achlioptas and Moore is, unfortunately for us, rather precise.  Recall
$\E[X^2]$ is bounded by above by $(n+1) \max_\alpha \left ( {2^{H(\alpha)} f_k(\alpha)^r}\right )^n$,
increasing $f_k(\alpha)$ by even an $\epsilon$-size constant increase 
$\E[X^2]$ exponentially, and the second moment bound is lost. 
We optimistically conjecture that the second moment method can be generalized
for $k > 3$, but it (currently) appears difficult;  it may require a different
approach to bounding $\E[X^2]$ rather than considering how $\beta(\sigma,\tau)$
depends on the overlap $\alpha$.  Possibly, the second moment method may not
be suitable for finding thresholds for general $k$, and other approaches (as in \cite{ding2015proof,ding2016satisfiability})
may be required. 

\section{Second Moment Argument for Hypergraph 2-Colorability}

\subsection{Adapting the Second Moment Argument}
We can similarly consider the second moment method with for hypergraph
2-colorability, another problem considered in
\cite{achlioptas2002,achlioptas2006random}.  
Here we follow the argument of \cite{achlioptas2002}.\footnote{In \cite{achlioptas2006random},
Achlioptas and Moore provided a more sophisticated 
argument with a simpler overall calculation, by considering {\em balanced} colorings, where the number of vertices of
each color are the same.  In our case, since the number of vertices is odd, we
cannot follow their argument completely, although it appears that the natural
approach of considering near-balanced colorings $\lfloor n/2 \rfloor$ vertices of the
first color and $\lceil n/2 \rceil$ vertices of the second color would allow
us to mimic their argument.  Because here we are using their mathematical derivations
as a black box, it suffices to use the approach of \cite{achlioptas2002}, which yields
the same result.}

Let $X$ be the number of 2-colorings of a hypergraph chosen from
${H}_k^{AP}(n,m)$, with $m=rn$ again.
We apply the second moment method as follows, for the case where $k=3$.
The probability that a 2-coloring with $z = \alpha n$ black vertices and
$n=z =(1-\alpha)n$ white vertices makes a random hyperedge bichromatic
is 
$$1-3\alpha+3\alpha^2 = 1 - \alpha^3 - (1-\alpha)^3,$$ 
based on Lemma~\ref{lem:mono}, just as it would be for a standard randomly hypergraph.
Thus 
$$\E[X] = \sum_{z=0}^n {n \choose z} (1 - (z/n)^3 - (1-z/n)^3)^{rn}.$$
Note that in this settings this expectation result cannot be generalized 
when $k > 3$, since as we have discussed the number monochromatic hyperedges would not depend only
on the fraction of vertices of each color.

To bound $\E[X^2]$, consider a pair of 2-colorings $S$ and $T$ with $\alpha n$
and $\beta n$ black vertices respectively, and $\gamma n$ vertices that are black
in both.  As described in \cite{achlioptas2002}, for a standard random hypergraph,
the probability a random hyperedge of size $k$ would be bichromatic under both 
$S$ and $T$ is given by:
$$1-\alpha^k - (1-\alpha)^k - \beta^k - (1-\beta)^k + \gamma^k + (\alpha-\gamma)^k 
+ (\beta-\gamma)^k  + (1-\alpha-\beta+\gamma)^k.$$ 
For the case of $k=3$, the above expression holds also for hypergraphs chosen from 
${H}_k^{AP}(n,m)$, again because of Lemma~\ref{lem:mono}. 
Let the expression above be denoted by $p(\alpha,\beta,\gamma)$.  
Achlioptas and Moore provide the following expression for $\E[X^2]$
in the setting of standard hypergraphs, which carries over here
as well:
$$\E[X^2] = \sum_{z_1,z_2,z_3,z_4} {n \choose z_1,z_2,z_3,z_4} p \left( \frac{z_1+z_2}{n}, \frac{z_1+z_3}{n}, 
\frac{z_3}{n} \right )^{rn},$$
where $z_1$ is the number of vertices colored black in both assignments, $z_2$ is
the number of vertices colored black in $S$ and white in $T$, 
$z_3$ is
the number of vertices colored white in $S$ and black in $T$, 
and $z_4$ is
the number of vertices colored white in both assignments.

Given these expression for $\E[X]$ and $\E[X^2]$, Achlioptas and Moore
show that the dominant when calculating the ratio between $\E[X^2]$ and
$\E[X]^2$ for the second moment method is when $\alpha = \beta = 1/2$
and $\gamma = 1/4$;  again, their calculations for the standard hypergraph
case carry over to the case of $k =3$ here, yielding the following final result.  

\begin{theorem}
For a random hypergraph ${H}_k^{AP}(n,m)$ with $m = rn$, $r < 3/2$, 
the hypergraph is 2-colorable with uniform positive probability.
\end{theorem}

\subsection{Remarks on the Result}

Again, we would like the with uniform positive probability result here
to yield a with high probability result.  The 2-colorability threshold
of standard random hypergraphs is considered explicitly by Friedgut
\cite{friedgut2005hunting}, who shows that there is a non-uniform
sharp threshold for hypergraph 2-colorability, which implies that in
the context of standard hypergraphs, the result with uniform positive
probability implies a with high probability result.  We conjecture
that the same holds here; however, the explicit proof by Friedgut
depends on a result by Erd\"{o}s and Simonovits on random $t$-partite
systems \cite{erdos1983supersaturated}.  It is an interesting question as to whether 
that result can generalize
to the setting of hyperedges governed by arithmetic progressions.

Also, as with \kapnae, it remains a challenge to obtain second moment
results for $k > 3$.  

\section{Conclusion}

Motivated by recent results in double hashing, we have introduced
quasi-random hypergraphs that we refer to as random AP hypergraphs.
Given that random AP hypergraphs have implicitly arisen in the study
of multiple-choice hashing data structures, they appear worthy of
future study.  We have conjectured that random AP hypergraphs share
the same thresholds as standard hypergrpahs with regard to
2-colorability, and have correspondingly defined AP versions of random
satisfiability problems, which we also conjecture share the same
satisfiability thresholds as their standard counterparts.  We have
shown that second moment arguments for thresholds for these problems
generalize when $k=3$.  We have also conjectured that the sharp/coarse
threshold results of Friedgut (and others) apply to random AP
hypergraphs as well.

At a higher level, it seems worthwhile to understand more clearly how
random AP hypergraphs are essentially similar to (and different from) standard
hypergraphs in generally usable ways.  Perhaps there is some
abstractable high-level property shared by random AP and standard
hypergraphs that would provide a mechanism for easily showing double
hashing versions of hashing data structures have the same performance
as standard versions, without an ad hoc analysis for each data
structure.  At the same time, it would be useful to understand how
random AP hypergraphs are different from standard hypergraphs; there
should be settings where the use of less randomness and/or specifically
the arithmetic progression constraint has a significant algorithmic or complexity-related effect.

\section*{Acknowledgments}

The author thanks Cris Moore for several helpful discussions, and Yufei Zhao for explaining results
regarding monochromatic colorings of arithmetic progressions of length 3.

\section*{Appendix}
\noindent {\bf Lemma 2.}
For a prime $n$, a two-coloring of $Z_n$ where the two color classes consist of $\alpha n$ numbers
and $(1-\alpha)n$ numbers has $(1-3\alpha+3\alpha^2)n^2$ monochromatic arithmetic sequence of length three.

\smallskip

\begin{proof}[Proof of Lemma 2]
Let $z_i$ be a variable for $0 \leq i < n$, $S$ be the set of all $n^2$ arithmetic progressions of length three given as ordered triples, and consider the polynomial
$$\sum_{(i,j,k) \in S} (z_iz_jz_k + (1-z_i)(1-z_j)(1-z_k)).$$
Now consider this polynomial for a given two-coloring of $Z_n$ with
$z_i$ being the color of $i \in Z_n$, and where one color class
corresponds to $z_i=0$ and the other corresponds to $z_i = 1$.  The
summation above then equals the number of monochromatic arithmetic
sequences, since each element in the sum is 1 if the triple $(i,j,k)$
corresponds to a monochromatic arithmetic sequence and 0 otherwise.
Expanding, we find
\begin{eqnarray*}
\sum_{(i,j,k) \in S} (z_iz_jz_k + (1-z_i)(1-z_j)(1-z_k)) & = & \sum_{(i,j,k) \in S} (1-z_i-z_j-z_k+z_iz_j+z_jz_k+z_iz_k) \\
& = & n^2 - 3n(\alpha n)+ 3(\alpha n) (\alpha n) \\
& = & (1 - 3\alpha + 3\alpha^2)n^2.  
\end{eqnarray*}
Note that for the second equality, we use that there are $n^2$ triples in $S$;  
there are $3n$ triples in $S$
including $i$ for the $\alpha n$ values of $z_i$ with $z_i = 1$ (counting the triple $(i,i,i)$ three times);  and there are 3 triples
in $S$ including $i$ and $j$ for every ordered pair $z_i$ and $z_j$ with $z_iz_j = 1$  (again counting the triple $(i,i,i)$ three times when $i=j$).  
\end{proof}

\end{document}